\documentclass{ws-procs9x6}
\newcommand{\cal}{}

\newcommand{\ub}{\bar u}
\newcommand\beq{\begin{eqnarray}}
\newcommand\eeq{\end{eqnarray}}
\newcommand\be{\begin{eqnarray}}
\newcommand\ee{\end{eqnarray}}

\def\Mom{{\cal M}}

\def\kslash{\slash{\mkern-1mu k}}

\def\ieps{i\varepsilon}
\def\pp{{\bf p}_\perp}
\def\kp{{\bf k}_\perp}

\begin{document}

\title{Violation of Sum Rules for Twist-3 Parton Distributions
             in QCD}

\author{M. BURKARDT}

\address{Dept. of Physics, New Mexico State University,
Las Cruces, NM 88003, USA\\
E-mail: burkardt@nmsu.edu}

\author{Y. KOIKE}

\address{Dept. of Physics, Niigata University,
Niigata 950--2181, Japan\\
E-mail: koike@nt.sc.niigata-u.ac.jp}


\maketitle

\abstracts{Sum rules for twist-3 distributions are reexamined.
Integral relations between twist-3 and twist-2
parton distributions suggest the possibility for
a $\delta$-function at $x=0$. We confirm and clarify
this result by constructing $h_L$ and $h_L^3$ (quark-gluon interaction
dependent part of $h_L$) 
explicitly from their moments for a one-loop
dressed massive quark.
The physics of these results is illustrated
by calculating $h_L(x,Q^2)$ using light-front 
time-ordered pQCD to ${O}(\alpha_S)$ on a 
quark target. 
}

\section{Introduction}
\setcounter{equation}{0}

Ongoing experiments with polarized beams and/or targets
conducted at RHIC, HERMES and COMPASS etc
are providing us with important information on the spin distribution
carried by quarks and gluons in the nucleon.
They are also enabling us to extract information on the
higher twist distributions which represent the effect of
quark-gluon correlations.  
In particular, the twist-3 distributions $g_T(x,Q^2)$ and $h_L(x,Q^2)$
are unique in that they appear as a leading contribution
in some spin asymmetries:  For example, $g_T$ can be measured in the 
transversely polarized lepton-nucleon deep inelastic scattering
and $h_L$ appears in the longitudinal-transverse spin asymmetry
in the polarized nucleon-nucleon Drell-Yan process\,\cite{JJ92}.
The purpose of this paper is to reexamine the
validity of the sum rules for these twist-3 distributions.

A complete list of twist-3 quark distributions
is given by the light-cone corelation functions in a
hadron with momentum $P$, spin $S$ and mass $M$:
\begin{eqnarray}
& &\int {d\lambda\over 2\pi}e^{i\lambda x}
\langle PS|\bar{\psi}(0)\gamma^\mu\gamma_5\psi(\lambda n)
|_{Q^2}|PS\rangle
\nonumber\\
& &\qquad=
2\left[g_1(x,Q^2)p^\mu (S\cdot n) 
+ g_T(x,Q^2)S_\perp^\mu +M^2 g_3(x,Q^2)n^\mu (S\cdot n)
\right],
\label{eq1}\\
& &\int{d\lambda\over 2\pi}e^{i\lambda x}
\langle PS | \bar{\psi}(0)\sigma^{\mu\nu}i\gamma_5 \psi(\lambda n)|_{Q^2}
|PS \rangle
=2\left[h_1(x,Q^2)(S_\perp^\mu p^\nu - S_\perp^\nu p^\mu )
/M\right.\nonumber\\
& &\left.\qquad
+h_L(x,Q^2)M(p^\mu n^\nu - p^\nu n^\mu )(S\cdot n)
+h_3(x,Q^2) M (S_\perp^\mu n^\nu - S_\perp^\nu n^\mu)\right],
\label{eq2}\\
& &\int{d\lambda\over 2\pi}e^{i\lambda x}
\langle PS|\bar{\psi}(0)\psi(\lambda n)|_{Q^2}|PS\rangle
=2Me(x,Q^2)\label{eq3}.
\end{eqnarray}
The light-like vectors $p$ and $n$
are introduced by the relation
$p^2=n^2=0$, $n^+=p^-=0$, $P^\mu = p^\mu +{M^2\over 2}n^\mu$ and
$S^\mu=(S\cdot n)p^\mu + (S\cdot p)n^\mu + S_\perp^\mu$.
The variable $x \in [-1,1]$ represents the parton's light-cone momentum fraction.  Anti-quark
distributions $\bar{g}_{1,T,3}(x,Q^2)$, $\bar{h}_{1,L,3}(x,Q^2)$ are
obtained by replacing $\psi$ by its charge
conjugation field $C\bar{\psi}^T$ in (\ref{eq1})-(\ref{eq3})
and are related to the quark distributions as
$\bar{g}_{1,T,3}(x,Q^2) = {g}_{1,T,3}(-x,Q^2)$ and
$\bar{h}_{1,T,3}(x,Q^2) = - {h}_{1,T,3}(-x,Q^2)$.  
The sum rules in our interest are obtained by taking the first
moment of the above relations.  For example, from (\ref{eq1}), one obtains
\begin{eqnarray}
& &\langle PS |\bar{\psi}(0)\gamma^\mu\gamma_5\psi(0)|_{Q^2}|PS\rangle
\label{eq4}\\
& &=2\int_{-1}^1\,dx\,\left[ g_1(x,Q^2)p^\mu(S\cdot n)+
g_T(x,Q^2)S_\perp^\mu + M^2 g_3(x,Q^2)n^\mu(S\cdot n)\right].
\nonumber
\end{eqnarray}
{}From rotational invariance, it follows that
the left hand side of (\ref{eq4}) is proportional to the spin vector
$S^\mu$ and thus $g_{1,T,3}(x,Q^2)$ must satisfy
\begin{eqnarray}
\int_{-1}^1\,dx\,g_1(x,Q^2) = \int_{-1}^1\,dx\,g_T(x,Q^2),
\label{eq5}\\
\int_{-1}^1\,dx\,g_1(x,Q^2) = 2\int_{-1}^1\,dx\,g_3(x,Q^2).
\label{eq6}
\end{eqnarray}
The same argument for (\ref{eq2}) leads to the sum rule relations
for $h_{1,L,3}(x,Q^2)$:
\begin{eqnarray}
\int_{-1}^1\,dx\,h_1(x,Q^2) = \int_{-1}^1\,dx\,h_L(x,Q^2),
\label{eq7}\\
\int_{-1}^1\,dx\,h_1(x,Q^2) = 2\int_{-1}^1\,dx\,h_3(x,Q^2).
\label{eq8}
\end{eqnarray}
The sum rule (\ref{eq5}) is known as 
Burkhardt-Cottingham sum rule\,\cite{BC}
and (\ref{eq7}) was first
derived in Refs. \cite{mb:nagoya,Bur95}.  
Since the twist-4 distributions $g_3$, $h_3$ are unlikely to be measured
experimentally,
the sum rules involving those functions
(\ref{eq6}) and (\ref{eq8}) are practically
useless and will not be addressed in the subsequent discussions.  
As is clear from the above derivation,
these sum rules are mere consequences of rotational invariance
and there is no doubt in their validity in a 
mathematical sense.
However, if one tries to confirm those sum rules by experiment,
great care is required to perform the integral including $x=0$.  
In DIS, $x$ is identified as the Bjorken
variable $x_B=Q^2/2P\cdot q$ and $x=0$ corresponds
to $P\cdot q\to \infty$ which can never be achieved
in a rigorous sense.  
Accordingly, if $h_L(x,Q^2)$ has 
a contribution proportional to $\delta(x)$ and $h_1(x,Q^2)$ does not,
experimental measurement would 
claim a violation of the sum rule.

In this paper we reexamine the sum rule 
involving the first moment of 
the twist-3 distribution 
$h_L(x,Q^2)$.  In particular we argue that
$h_L(x,Q^2)$ has a $\delta$-singularity
at $x=0$. 
Starting from the general QCD-based 
decomposition of $h_L$, we show that it 
contains a
function $h_L^m$ which has a $\delta(x)$-singularity.  
In Sec.3, we construct $h_L$ for a massive 
quark from the moments of $h_L^3$ at the 
one-loop level
and show that $h_L^3$ also has an 
$\delta(x)$-singularity,
which together with the singularity in $h_L^m$ gives 
rise to a $\delta(x)$ singularity in $h_L$ itself. 
In Sec.4, we perform an explicit light-cone 
calculation of $h_L$ in the one-loop level to confirm the 
result of the previous sections.
Details can be found in Ref. 
\cite{us}.

\section{$\delta(x)$-functions in $h_L(x,Q^2)$}
\setcounter{equation}{0}
The OPE analysis of the correlation function (\ref{eq2}) allows us to
decompose $h_L(x,Q^2)$ into the contribution 
expressed in terms of twist-2 distributions
and the rest which we call $h_L^3(x,Q^2)$.
Since the scale dependence of each distribution
is inessential in the following discussion, we shall omit it in this 
section
for simplicity.  
Introducing the notation for the moments on $[-1,1]$,
$
{\cal M}_n[h_L]\equiv \int_{-1}^1\,dx\,x^nh_L(x),$
this decomposition is given in terms of the moment relation\,\cite{JJ92}:
\beq
\Mom_n[h_L] &=& {2\over n+2}\Mom_n[h_1] + {n\over n+2}{m_q\over M}\Mom_{n-1}[g_1]
+\Mom_n[h_L^3],\qquad (n\geq 1)
\label{eq:hlmomn}\\
\Mom_0[h_L] &=& \Mom_0[h_1],
\label{eq:hlmom0}
\eeq
with the conditions 
\beq
\Mom_0[h_L^3]=
\Mom_1[h_L^3]=0.
\label{eq:hl3mom1}
\eeq
By inverting the moment relation, one finds
\beq
h_L(x) &=& h_L^{WW}(x) + h_L^m(x) + h_L^3(x)
\label{eq:ope1a}\\
& & \!\!\!\!\!\!\!\!\! \!\!\!\!\!\!\!\!\!\!
\!\!\!\!\!
=\left\{
\begin{array}{l}
\displaystyle{
2x\int_x^1dy \frac{h_1 (y)}{y^2}
+\frac{m_q}{M} \left[ \frac{g_1(x)}{x} -2x \int_x^1 dy
  \frac{g_1(y)}{y^3}\right] + h_L^3(x) }
\qquad\qquad (x>0) \\[12pt] 
\displaystyle{
 -2x\int_{-1}^xdy \frac{h_1 (y)}{y^2}
+\frac{m_q}{M} \left[ \frac{g_1(x)}{x} +2x \int_{-1}^x dy
  \frac{g_1(y)}{y^3}\right] + h_L^3(x),
} 
\qquad \,\,\,(x<0) 
\end{array}
\right.
\label{eq:ope1b}
\eeq
where the first and second terms in Eq. 
(\ref{eq:ope1a}) denote the corresponding terms 
in (\ref{eq:ope1b}).
In this notation the sum rule (\ref{eq:hlmom0}) and the condition
(\ref{eq:hl3mom1}) implies\,
\footnote{More precisely, the original OPE tells us
${\cal M}_0[h_L^3+h_L^m]=0$.   But as long as $g_1(0_\pm)$ is finite,
which we will assume, this is equivalent to
stronger relations (\ref{eq:hl3mom1}) and 
(\ref{eq:hlm0}).}
\beq
{\cal M}_0[h_L^m]=0.
\label{eq:hlm0}
\eeq
Naively integrating 
(\ref{eq:ope1b}) 
over $x$ for $x>0$, 
while dropping all surface terms \cite{daniel} one
arrives at $\int_0^1dx h_L(x)=\int_0^1 dx h_1(x)
+\int_0^1 dx h_L^3(x)$ and likewise for $\int_{-1}^0 dxh_L(x)$.
Together with (\ref{eq:hl3mom1}), this
yields
$\int_{-1}^1 dx h_L=\int_{-1}^1 dx h_1$.
However, this procedure
may be wrong due to the very
singular behavior of the functions involved
near $x=0$. Investigating this issue
will be the main purpose of this paper.

We first address the potential singularity at $x=0$ in the integral
expression for $h_L^m(x)$ in (\ref{eq:ope1b}).    
In order to regulate the region near $x=0$, we first
multiply $h_L^m(x)$
by $x^\beta$, integrate from $0$ to $1$
and let $\beta \rightarrow 0$. This yields
\beq
\int_{0+}^1 dx h_L^m(x) 
= \frac{m_q}{2M}\lim_{\beta \rightarrow 0} \beta \int_0^1 dx x^{\beta -1}
g_1(y)
={m_q\over 2M} g_1(0+), 
\label{eq:ope2a}
\eeq 
while multiplying Eq. (\ref{eq:ope1b}) by $|x|^\beta$ and integration from
$-1$ to $0$ yields
\beq
\int_{-1}^{0-} dx h_L^m(x)
= -\frac{m_q}{2M}\lim_{\beta \rightarrow 0} \beta \int_{-1}^{0-} dx 
|x|^{\beta -1}
g_1(y)
=-{m_q\over 2M} g_1(0-),
\label{eq:ope2b}
\eeq 
where we have assumed that $g_1(0\pm)$ is finite.  
Adding these results we have
\beq
\int_{-1}^{0-}dx h_L^m(x)+ \int_{0+}^1dx h_L^m(x) =
{m_q\over 2M} \left(g_1(0+)-g_1(0-)\right).
\label{eq:hlmsing}
\eeq
Eq. (\ref{eq:hlm0})
and the fact that, in general, 
$\lim_{x\rightarrow 0}
g_1(x) - g_1(-x)\neq 0$, \footnote{For example,
dressing a quark at ${\cal O}(\alpha_S)$ yields
$g_1(0+)\neq 0$ and $g_1(0-)\equiv \bar{g}_1(0+)=0$.}
imply
\beq
h_L^m(x)=h_L^m(x)_{reg} -{m_q\over 2M} \left(g_1(0+)-g_1(0-)\right)
\delta(x),
\label{eq:hlmreg}
\eeq
where $h_L^m(x)_{reg}$ is defined 
by the integral in (\ref{eq:ope1b}) at $x>0$ and $x<0$ and is regular
at $x=0$.  
Eq. (\ref{eq:hlmreg}) indicates
that $h_L$ has a $\delta(x)$ term unless $h_L^3(x)$ has a 
$\delta(x)$ term and it cancels the above 
singularity in $h_L^m(x)$.  

Eq. (\ref{eq:hlmreg}) demonstrates that 
the functions constituting $h_L(x)$ 
are more singular
near $x=0$ than previously assumed and great
care needs to be taken when replacing integrals
over nonzero values of $x$ by integrals that 
involve the origin. 
In particular,
if $h_L(x)$
itself contains a $\delta(x)$ term, then 
(\ref{eq:hlmom0}) implies
\beq
\int_{0+}^1\left(h_L(x)-h_1(x)\right)
+\int_{-1}^{0-}\left(h_L(x)-h_1(x)\right)
\neq 0,
\eeq
and, since $h_1(x)$ is singularity free at $x=0$:
\beq
\int_{-1}^{0-}dx h_1(x)+ \int_{0+}^1dx h_1(x) = 
\int_{-1}^1 dx h_1(x).
\label{eq:h10}
\eeq
Accordingly an attempt to verify the ``$h_L$-sum 
rule'' \cite{mb:nagoya}
would obviously fail.

However, 
in order to see whether the 
$\delta(x)$ identified in (\ref{eq:hlmreg}) 
eventually survives in 
$h_L(x)$, we have to investigate the behavior of $h_L^3(x)$ at $x=0$.  
To this end 
we will explicitly construct $h_L(x)$ for a massive
quark to ${\cal O}(\alpha_S)$.

\section{$h_L(x,Q^2)$ from the moment relations}
In this section we will construct 
$h_L(x,Q^2)$ for a massive quark (mass $m_q$)
to ${\cal O}(\alpha_S)$
{}from the one-loop calculation of ${\cal M}_n[h_L^3]$.   
\beq
h_L(x,Q^2)=h_L^{(0)}(x) + {\alpha_S\over 2\pi}C_F{\rm ln}{Q^2\over m_q^2}
h_L^{(1)}(x),
\label{eq:hl(1)}
\eeq
where the scale $Q^2$ is introduced as an ultraviolet cutoff and
the $C_F=4/3$ is the color factor. 
$h_L^{WW,3,m(0,1)}(x)$ are defined similarly.  
$g_1^{(0)}(x)=h_1^{(0)}(x)=\delta (1-x)$ gives 
$h_L^{(0)}(x)=\delta(1-x)$, as it should.
One loop calculations for $g_1(x)$ and $h_1(x)$ 
for a quark yield
the well known splitting functions\,\cite{DGLAP,AM90}:
\beq
g_1^{(1)}(x)&=&{1+x^2 \over [1-x]_+} + 
{3\over 2}\delta(1-x),
\label{eq:g1(1)}\\
h_1^{(1)}(x)&=& {2x\over [1-x]_+} + {3\over 2}\delta(1-x).
\label{eq:h1(1)}  
\eeq
Inserting these equations into the defining equation in
(\ref{eq:ope1b}), one obtains
\beq
& &\!\!\!\!
h_L^{WW(1)}(x)=3x + 4x{\rm ln} {1-x\over x},
\label{eq:hlww(1)}\\
& &\!\!\!\!h_L^{m(1)}(x)=
{2\over (1-x)_+}-4x{\rm ln}{1-x\over x}-3 +3x
-{1\over 2}\delta(x).
\label{eq:hlm(1)}
\eeq
In the first line of (\ref{eq:hlm(1)}), 
the term $(3x-{3\over 2}\delta(1-x))$ comes from
the self-energy correction, i.e. from expanding
$M=m_q\left[1+\frac{\alpha_S}{2\pi}C_F \frac{3}{2}\ln 
\frac{Q^2}{m_q^2}\right]$ in Eq. (\ref{eq:ope1a}), 
and $-\frac{1}{2}\delta(x)=-\frac{1}{2}g_1(0+)\delta(x)$ 
in $h_L^{m(1)}(x)$ accounts for the second term on 
the r.h.s. of (\ref{eq:hlmreg}).  
We also note that $h_L^{WW(1)}$ does not have any 
singularity at $x=0$ and satisfies
$\int_{0}^1\,dx\,h_L^{WW(1)}(x) =
\int_{0}^1\,dx\,h_1^{(1)}(x)$ as it should.

$h_L^{(1)}(x)$ can be constructed if we know 
the purely twist-3 part $h_L^{3(1)}(x)$ at 
the one-loop level.   
One-loop renormalization of $h_L$ was completed in \cite{KT95}
and the mixing matrix for the local operators
contributing to the moments of $h_L(x,Q^2)$ was presented.  
We obtain for the
moment of the quark distribution\cite{us}
\beq
\int_{-1}^1\,dx\,x^nh_L^{3(1)}(x) 
={3\over n+1} -{6\over n+2} +{1\over 2} 
\label{eq:hl3(1)mom}
\eeq
for $n\geq 2$.
{}From this result, together with (\ref{eq:hl3mom1}),
we can construct $h_L^{3(1)}(x)$ as 
\beq
h_L^{3(1)}(x)= 3 -6x + {1\over 2}\delta(1-x) -{1\over 2}\delta(x)
\label{eq:hl3(1)}.  
\eeq
We emphasize that
the $-1/2\delta(x)$ in (\ref{eq:hl3(1)}) is necessary to reproduce the 
$n=0$ moment of $h_L^{3(1)}(x)$. From (\ref{eq:hlww(1)}), 
(\ref{eq:hlm(1)}) and (\ref{eq:hl3(1)}), one obtains
\beq
h_L(x,Q^2) = \delta(1-x) +
{\alpha_S\over 2\pi}{\rm ln}{Q^2\over m_q^2}C_F
\left[{2\over [1-x]_+} +{1\over 2}\delta(1-x) -\delta(x)\right].
\label{eq:hlx}
\eeq
We remark that
the above calculation indicates
that the $\delta(x)$ term appears not only in
$h_L^{m}$ but also in $h_L^{3}$.  
Furthermore they do not cancel but add up to
give rise to $-\delta(x)$ in $h_L(x,Q^2)$ itself.  
In the next section we will confirm Eq. 
(\ref{eq:hlx}) through
a direct calculation of $h_L(x,Q^2)$ for a quark.

\section{Light-cone calculation of $h_L(x,Q^2)$}
In order to illustrate the physical origin of the $\delta(x)$ terms in
$h_L(x)$ and to develop a more convenient procedure
for calculating such terms, we now evaluate $h_L(x)$ using
time-ordered light-front (LF) perturbation theory.
The method has been outlined in Ref. \cite{HZ} and we
will restrict ourselves here to the essential steps only.
There are two equivalent ways to perform time-ordered LF perturbation
theory: one can either work with the LF Hamiltonian for QCD and perform
old-fashioned perturbation theory\cite{HZ}, 
or one can start from Feynman perturbation theory and integrate over the
LF-energy $k^-$ first. In the following, we will use
the latter approach for
the one-loop calculation of $h_L(x)$.

In LF gauge, $A^+=0$, parton distributions can be expressed
in terms of LF momentum densities
($k^+$-densities). Therefore, one finds for a parton distribution,
characterized by the Dirac matrix $\Gamma$
at ${\cal O}(\alpha_S)$ and for $0<x<1$
\beq
f_\Gamma(x) \ub(p)\Gamma u(p)&=& -ig^2 \ub(p)
\int \frac{d^4k}{(2\pi)^4}
\gamma^\mu \frac{1}{\kslash-m_q+\ieps}
\Gamma \frac{1}{\kslash-m_q+\ieps}\gamma^\nu u(p)
\nonumber\\ & &\qquad \qquad \qquad \times
\delta\left(x-\frac{k^+}{p^+}\right)
D_{\mu \nu}(p-k),
\label{eq:lf1}
\eeq
where
$
D_{\mu \nu}(q)
=\frac{1}{q^2+\ieps}\left[
g_{\mu \nu}-\frac{q_\mu n_\nu+n_\mu q_\nu}{qn}
\right]
$
is the gauge field propagator in LF gauge, and
$n^\mu$ is a light-like vector such that
$nA=A^+ \sim\left(A^0+A^3\right)/\sqrt{2}$ for any
four vector $A^\mu$.
The $k^-$ integrals in expressions like Eq.
(\ref{eq:lf1}) are performed using Cauchy's theorem,
yielding for $0<k^+<p^+$
\beq
&-i&\int \frac{dk^-}{2\pi}
\frac{1}{\left(k^2-m_q^2+\ieps\right)^2}
\frac{1}{(p-k)^2+\ieps}\\
&=& \frac{1}{(2k^+)^2}\frac{1}{2(p^+-k^+)}
\frac{1}{\left(p^--\frac{m_q^2+\kp^2}{
2k^+} - \frac{ (\pp-\kp)^2}{2(p^+-k^+)}
\right)^2}
\stackrel{\kp \rightarrow \infty}
{\longrightarrow}
\frac{1}{2p^+} \frac{1-x}{\kp^4},\nonumber
\eeq
where we used $k^+=xp^+$.
In order to integrate all terms in Eq. (\ref{eq:lf1}) over $k^-$,
Cauchy's theorem is used to replace any factors of 
$k^-$ in the numerator of Eq. (\ref{eq:lf1}) 
containing $k^-$ by their on-shell value at the pole 
of the gluon propagator
\beq
k^- \longrightarrow \tilde{k}^- \equiv
p^--\frac{(\pp-\kp)^2}{2(p^+-k^+)}
\label{eq:ktilde}
\eeq
In the following we will focus on the UV divergent 
contributions to the parton distribution only. This 
helps to keep the necessary algebra at a reasonable 
level.
We find for $0<x<1$ to ${\cal O}(\alpha_S)$
\beq
h_L(x,Q^2)= \frac{\alpha_S}{2\pi}C_F 
\ln \frac{Q^2}{m_q^2} \frac{2}{[1-x]_+},
\label{eq:hL1}
\eeq
where the usual $+$-prescription for $\frac{1}{[1-x]_+}$ applies at
$x=1$, i.e.  $\frac{1}{[1-x]_+}=\frac{1}{1-x}$ for $x<1$ and
$\int_0^1dx\frac{1}{[1-x]_+}=0$.
Furthermore, $h_L(x)=0$ for $x<0$, since anti-quarks do not occur in the 
${\cal O}(\alpha_S)$ dressing of a quark. In addition to Eq. (\ref{eq:hL1}), 
there is also an explicit $\delta(x-1)$ contribution at $x=1$. These
are familiar from twist-2 distributions, where they reflect the fact
that the probability to find the quark as a bare quark is less than one
due to the dressing with gluons. For higher-twist
distributions, the wave function renormalization contributes is 
$\frac{\alpha_S}{2\pi}C_F \ln \frac{Q^2}{m_q^2}
\frac{3}{2}\delta(x-1)$. The same wave function renormalization
also contributes at twist-3.
However, for all higher twist distributions there is an additional
source for $\delta(x-1)$ terms
which has, in parton language, more the appearance
of a vertex correction, 
but which
arises in fact from the gauge-piece of self-energies connected to the
vertex by an `instantaneous fermion propagator' $\frac{\gamma^+}{2p^+}$.
For $g_T(x,Q^2)$ these have been calculated in Ref.
\cite{HZ} where they give an additional
contribution $-\frac{\alpha_S}{2\pi}C_F \ln \frac{Q^2}{m_q^2}
\delta(x-1)$, i.e. the total contribution at $x=1$ for $g_T(x,Q^2)$ was
found to be $\frac{\alpha_S}{2\pi}C_F \ln \frac{Q^2}{m_q^2}
\frac{1}{2}\delta(x-1)$. We found the same $\delta(x-1)$ terms also
for $h_L(x,Q^2)$.
\footnote{In LF gauge, different components of 
the fermion field aquire different wave function 
renormalization. However, since all twist-3 parton
distributions involve one LC-good and one LC-bad 
component, one finds the same wave 
function renormalization for all three twist-3 
distributions.}
Combining the $\delta(x-1)$ piece with Eq. 
(\ref{eq:hL1}) we thus find for $0<x<1$
\beq
h_L(x,Q^2) = \delta(x-1) +
\frac{\alpha_S}{2\pi}
C_F\ln\frac{Q^2}{m_q^2}\left[\frac{2}{[1-x]_+} + 
\frac{1}{2}\delta(x-1)\right]
.\label{eq:hlcan}
\eeq
Comparing this result with the well known result for 
$h_1$\,\cite{AM90}
\beq
h_1(x,Q^2) = \delta(x-1)
+ \frac{\alpha_S}{2\pi} \ln \frac{Q^2}{m_q^2} C_F
\left[\frac{2x}{[1-x]_+} + 
\frac{3}{2}\delta(x-1)\right],
\eeq
one realizes that
\beq
\lim_{\varepsilon \rightarrow 0}
\int_\varepsilon^1 dx \left[
h_L(x,Q^2)-h_1(x,Q^2\right]
=   \frac{\alpha_S}{2\pi} \ln \frac{Q^2}{m_q^2} C_F
\neq 0,
\eeq
i.e. if one excludes the possibly
problematic region $x=0$, then the $h_L$-sum rule
\cite{mb:nagoya} is 
violated already for a quark dressed with gluons 
at order ${\cal O}(\alpha_S)$.

In the above calculation, we carefully avoided
the point $x=0$.
For most values of $k^+$, the denominator in
(\ref{eq:lf1}) contains three powers of 
$k^-$ when $k^-\rightarrow \infty$.
However, when $k^+=0$, $k^2-m_q^2$ becomes
independent of $k^-$ and the denominator in
(\ref{eq:lf1}) contains only one power of
$k^-$. Therefore, for those terms in the numerator
which are linear in $k^-$, 
the $k^-$-integral diverges linearly.
Although this happens only for a point of measure
zero (namely at $k^+=0$), a linear divergence is indicative
of a singularity of $h_L(x,Q^2)$ at that point.
\footnote{Note that the divergence at 
$k^+=p^+$ is only logarithmic.}
To investigate the $k^+\approx 0$ singularity in
these terms further, we consider
\beq
f(k^+,\kp)&\equiv& \int\!\! dk^-\!
\frac{k^-}{(k^2-m_q^2+\ieps)^2}
\frac{1}{(p-k)^2+\ieps}
\label{eq:fsing1}\\
& &\!\!\!\!\!\!\!\!\!\!\!\!\!\!\!\!\!\!\!=
\int\!\! dk^-\!
\frac{\tilde{k}^-\quad+\quad 
\left(k^--\tilde{k}^-\right)}
{(k^2-m_q^2+\ieps)^2\left[(p-k)^2+\ieps\right]} 
=f_{can.}( k^+,\kp) + f_{sin.}(k^+,\kp),
\nonumber
\eeq
where the `canonical' piece $f_{can.}$
is obtained by
substituting for $k^-$ its on energy-shell value 
$\tilde{k}^-= p^- -\frac{(\pp-\kp)^2}{2(p^+-k^+)}$
[the value at the pole at $(p-k)^2=0$
, Eq. (\ref{eq:ktilde})].
For $k^+=xp^+\neq 0$, it is only this canonical
piece which contributes. To see this, we note that
$
k^--\tilde{k^-} = - \frac{(p-k)^2}{2(p^+-k^+)},
$
and therefore
\beq
f_{sin}(k^+,\kp)&=& \int dk^- \frac{k^--\tilde{k}^-}{(k^2-m_q^2+\ieps)^2}
\frac{1}{(p-k)^2+\ieps}
\nonumber\\
&=& \frac{1}{2(p^+-k^+)}\int dk^- \frac{1}{(k^2-m_q^2+\ieps)^2}
\label{eq:sin}.
\eeq
Obviously\cite{CY}
$
\int dk^- \frac{1}{(2k^+k^--\kp^2-m_q^2+\ieps)^2}=0
$
for $k^+\neq 0$ because then one can always avoid
enclosing the pole at $k^-=\frac{m_q^2+\kp^2-\ieps}
{2k^+}$ by closing the contour in the appropriate 
half-plane of the complex $k^--plane$.
However, on the other hand
$
\int \!\!d^2k_L \!\frac{1}{(k_L^2-\kp^2-m_q^2+\ieps)^2}
=\frac{i\pi}{\kp^2+m_q^2}
$
and therefore
\beq
f_{sin}(k^+,\kp) = {1\over 2p^+} \frac{i\pi\delta(k^+)}{\kp^2+m_q^2}.
\label{eq:delta}
\eeq
Upon collecting all terms $\propto k^-$ in the
numerator of (\ref{eq:lf1}), and applying 
(\ref{eq:delta}) to those terms we find 
for the
terms in $h_L(x,Q^2)$ that are singular at $x=0$
\beq
h_{L,sin}(x,Q^2)=
-\frac{\alpha_S}{2\pi} \ln \frac{Q^2}{m_q^2} C_F \delta (x).
\eeq
Together with Eq. (\ref{eq:hlcan}), this gives our 
final result for $h_L$, up to ${\cal O}(\alpha_S)$, 
valid also for $x=0$
\beq
h_L(x,Q^2) = \delta(x-1) +
\frac{\alpha_S}{2\pi}
C_F\ln\frac{Q^2}{m_q^2}\left[-\delta(x)+
\frac{2}{[1-x]_+} + 
\frac{1}{2}\delta(x-1)\right]
.\label{eq:hl}
\eeq
As expected, $h_L$ from Eq. (\ref{eq:hl}) does now satisfy
the $h_L$-sum rule, provided of course the origin is included 
in the integration.

This result is important for several reasons. First of all
it confirms our result for $h_L(x,Q^2)$ as determined from the
moment relations. Secondly, it provides us with a 
method for
calculating these $\delta(x)$ terms and thus
enabling us to address the issue of validity of the naive
sum rules more systematically. And finally, it shows that
there is a close relationship between these $\delta(x)$ terms
and the infamous zero-modes in LF field theory \cite{mb:adv}.

Ref. \cite{metz}, where canonical Hamiltonian
light-cone perturbation theory is used to 
calculate $h_L(x)$ and,
for $x\neq 0$ the result obtained in Ref. 
\cite{metz} agrees
with ours which provides an independent check of the formalism 
and the algebra.  However, the canonical light-cone perturbation
theory used in Ref. \cite{metz} is not adequate for studying the
point $x=0$.  From the smooth behaviour of $h_L(x)$ {\it near} 
$x=0$ the authors of Ref. \cite{metz} conclude that the sum
rule for the parton distribution $h_L(x)$ is violated to
${\cal O}(\alpha_S)$.  Our explicit calculation 
for $h_L(x)$ not only proves that the sum rule for $h_L(x)$
is {\it not} violated to this order if the point $x=0$ is properly 
included, but also shows that it is incorrect to draw conclusions 
{}from smooth behaviout near $x=0$ about the behaviour at $x=0$.

\section{Summary}
\setcounter{equation}{0}
We have investigated the
twist-3 distribution $h_L(x)$,
and found that the sum-rule for its lowest moment
is violated if the point $x=0$ is not properly included.

For a massive quark, to ${\cal O}(\alpha_S)$ we found
\beq
h_L(x,Q^2) &=& \delta(x-1) +
\frac{\alpha_S}{2\pi}
C_F\ln\frac{Q^2}{m_q^2}\left[-\delta(x)+\frac{2}{[1-x]_+} + 
\frac{1}{2}\delta(x-1)\right]
.\label{eq:higherT}
\eeq
At ${\cal O}(\alpha_S)$, $h_L(x,Q^2)$ does not 
satisfy its
sum rule if one excludes the origin from
the region of integration 
(which normally happens in 
experimental attempts to verify a sum rule). 
Of course, QCD is a strongly interacting theory and parton
distribution functions in QCD are nonperturbative observables.
Nevertheless, if one can show that a sum rule fails already
in perturbation theory, then this is usually a very strong
indication that the sum rule also fails nonperturbatively
(while the converse is often not the case!).

{}From the QCD equations of motion, we were able to show 
nonperturbatively that
the difference between $h_L(x,Q^2)$ and $h_L^3(x,Q^2)$ contains
a $\delta(x)$ term
\beq
\left[h_L(x,Q^2)-h_L^3(x,Q^2)\right]_{singular}=
-\frac{m_q}{2M}\left(g_1(0+,Q^2)-g_1(0-,Q^2)\right)\delta(x).
\eeq
Since $g_1(0+,Q^2)-g_1(0-,Q^2)\equiv \lim_{x\rightarrow 0}
g_1(x,Q^2)-\bar{g}_1(x.Q^2)$ seems to be nonzero (it may even 
diverge\footnote{
In the next-to-leading order QCD for a quark, 
$\lim_{x\to 0}g_1(x)-\bar{g}_1(x)$ is 
logarithmically divergent\,\cite{g1nlo}.}), 
one can thus conclude that either $h_L(x,Q^2)$ or
$h_L^3(x,Q^2)$ or both do contain such a singular term.

We checked the validity of this relation to ${\cal O}(\alpha_S)$
and found that, to this order, both $h_L^3$ and $h_L$
contain a term $\propto \delta(x)$. We also verified that
even though the sum rule for $h_L(x)$ is violated if
$x=0$ is not included, it is still satisfied to ${\cal O}(\alpha_S)$
if the contribution from $x=0$ (the $\delta(x)$ term) is included.

\section*{Acknowledgments}
M.B. was supported by a grant from the DOE (FG03-95ER40965).  Y.K. is supported by the Grant-in-Aid for Scientific
Research (No. 12640260) of the Ministry of Education, Culture, Sports,
Science and Technology (Japan).    
We are also greatful to JSPS for the Invitation Fellowship
for Research in Japan (S-00209) which made it possible to 
materialize this work.

\end{document}